\documentstyle[]{mn}% If your system has the AMS fonts version 2.0 installed, MN.sty can be
\newif\ifAMStwofonts
\ifoldfss
  \newcommand{\rmn}[1] {{\rm #1}}

  \ifCUPmtlplainloaded \else
    \NewTextAlphabet{textbfit} {cmbxti10} {}
    \NewTextAlphabet{textbfss} {cmssbx10} {}
    \NewMathAlphabet{mathbfit} {cmbxti10} {} % for math mode
    \NewMathAlphabet{mathbfss} {cmssbx10} {} %  "   "    "
  \fi
  \ifAMStwofonts
    \ifCUPmtlplainloaded \else
      \NewSymbolFont{upmath} {eurm10}
      \NewSymbolFont{AMSa} {msam10}
      \NewMathSymbol{\upi}     {0}{upmath}{19}
      \NewMathSymbol{\umu}     {0}{upmath}{16}
      \NewMathSymbol{\upartial}{0}{upmath}{40}
      \NewMathSymbol{\leqslant}{3}{AMSa}{36}
      \NewMathSymbol{\geqslant}{3}{AMSa}{3E}

      \let\leq=\leqslant 
      \let\geq=\geqslant 
    \fi
  \fi
\fi % End of OFSS

\ifnfssone
  \newmathalphabet{\mathit}
  \addtoversion{normal}{\mathit}{cmr}{m}{it}
  \addtoversion{bold}{\mathit}{cmr}{bx}{it}
  \newcommand{\rmn}[1] {\mathrm{#1}}

  \newmathalphabet{\mathbfit} % math mode version of \textbfit{..}
  \addtoversion{normal}{\mathbfit}{cmr}{bx}{it}
  \addtoversion{bold}{\mathbfit}{cmr}{bx}{it}
  \newmathalphabet{\mathbfss} % math mode version of \textbfss{..}
  \addtoversion{normal}{\mathbfss}{cmss}{bx}{n}
  \addtoversion{bold}{\mathbfss}{cmss}{bx}{n}
  \ifAMStwofonts
    \ifCUPmtlplainloaded \else
      \UseAMStwoboldmath
      \makeatletter
      \new@mathgroup\upmath@group
      \define@mathgroup\mv@normal\upmath@group{eur}{m}{n}
      \define@mathgroup\mv@bold\upmath@group{eur}{b}{n}
      \edef\UPM{\hexnumber\upmath@group}
      \new@mathgroup\amsa@group
      \define@mathgroup\mv@normal\amsa@group{msa}{m}{n}
      \define@mathgroup\mv@bold\amsa@group{msa}{m}{n}
      \edef\AMSa{\hexnumber\amsa@group}
      \makeatother
      \mathchardef\upi="0\UPM19
      \mathchardef\umu="0\UPM16
      \mathchardef\upartial="0\UPM40
      \mathchardef\leqslant="3\AMSa36
      \mathchardef\geqslant="3\AMSa3E

      \let\leq=\leqslant 
      \let\geq=\geqslant 
    \fi
  \fi
\fi % End of NFSS release 1

\ifnfsstwo
  \newcommand{\rmn}[1] {\mathrm{#1}}

  \DeclareMathAlphabet{\mathbfit}{OT1}{cmr}{bx}{it}
  \SetMathAlphabet\mathbfit{bold}{OT1}{cmr}{bx}{it}
  \DeclareMathAlphabet{\mathbfss}{OT1}{cmss}{bx}{n}
  \SetMathAlphabet\mathbfss{bold}{OT1}{cmss}{bx}{n}
  \ifAMStwofonts
    \ifCUPmtlplainloaded \else
      \DeclareSymbolFont{UPM}{U}{eur}{m}{n}
      \SetSymbolFont{UPM}{bold}{U}{eur}{b}{n}
      \DeclareSymbolFont{AMSa}{U}{msa}{m}{n}
      \DeclareMathSymbol{\upi}{0}{UPM}{"19}
      \DeclareMathSymbol{\umu}{0}{UPM}{"16}
      \DeclareMathSymbol{\upartial}{0}{UPM}{"40}
      \DeclareMathSymbol{\leqslant}{3}{AMSa}{"36}
      \DeclareMathSymbol{\geqslant}{3}{AMSa}{"3E}

      \let\leq=\leqslant 
      \let\geq=\geqslant 
    \fi
  \fi
\fi % End of NFSS release 2

\ifCUPmtlplainloaded \else
  \ifAMStwofonts \else % If no AMS fonts
    \def\upi{\pi}
    \def\umu{\mu}
    \def\upartial{\partial}
  \fi
\fi

\title[Giant low surface brightness halos in distant radio galaxies]
{Giant low surface brightness halos  in distant radio galaxies: USS0828+193}
\author[Villar-Mart\'\i n et al.]{M. Villar-Mart\'\i n$^1$, J. Vernet$^2$,  
S. di Serego Alighieri$^2$, R. Fosbury$^3$, 
L. Pentericci $^4$, 
\newauthor M. Cohen$^5$, R. Goodrich$^6$, A. Humphrey$^1$\\
$^1$Dept. of Physical Sciences, University of Hertfordshire, College Lane, Hatfield,
Herts, AL10 9AB, UK\\
$^2$Osservatorio Astrofisico di Arcetri, Largo E. Fermi 5, I-50125, Firenze,
 Italy\\
$^3$Space Telecope European Coordinating Facitily,
Karl Schwarschild Str. 2, D-85748 Garching, bei Muenchen, Germany\\
$^4$Max Plank Institute fur Astronomie, Konigstuhl 17, D-69117 Heidelberg, Germany\\
$^5$California Institute of Technology, Mail Stop 105-24, Pasadena, CA 91125, USA\\
$^6$W.M. Keck Observatory 65-1120 Mamalahoa Highway, Kamuela, HI 96742, USA\\}

\date{}

\pagerange{\pageref{firstpage}--\pageref{lastpage}}
\pubyear{2001}

\begin{document}

\maketitle

\label{firstpage}

\begin{abstract}

	We present results on the spectroscopic study of the ionized gas in
the high redshift radio galaxy USS0828+193 at $z=$2.57. Thanks to the high 
S/N of the emission lines in the Keck spectrum, we have been able to perform
a detailed kinematic study by means of the spectral decomposition of the emission line
profiles.
This study reveals the existence of two types of material in this object: a) a
 low surface brightness component with   apparent quiescent kinematics consistent with
gravitational motions and b) a  perturbed component with   rather extreme kinematics.
The quiescent halo extends across the entire object for $\sim$80 kpc. It is enriched with heavy 
elements and 
apparently ionized by the continuum 
from the active nucleus.

	The properties of the quiescent halo and its origin are discussed in this paper.
We propose that it could be part of a structure that surrounds the entire object,
although its nature is not clear (a rotating disc? low surface brightness satellites? a cooling flow
nebula? material ejected in galactic winds? other?).

\end{abstract}


\section{Introduction}

	Extended Ly$\alpha$ regions are a common feature of high redshift 
radio galaxies ($z>$2, HzRG)
and quasars (Heckman et
al. 1991; see also narrow band Ly$\alpha$ images of HzRG in, e.g, Kurk et al. 2001;
Chambers, Miley \& van Breugel  1990; McCarthy et al. 1990b). 
Most morphological and kinematic 
studies are based on the high surface brightness regions. 
These regions are clumpy, irregular and often aligned with the radio axis.
They are characterized by extreme kinema\-tics, with measured 
FWHM and velocity shifts  $\geq$1000 km s$^{-1}$ 
(Baum \& McCarthy 2000,  Villar-Mart\'\i n, Binette \& Fosbury 
1999, McCarthy, Baum \& Spinrad 1996), compared to values
of $\sim$few hundreds in low redshift radio galaxies (Baum, Heckman \& van Breugel 1990,
Tadhunter, Fosbury \& Quinn 1989). Although gravitational motions
cannot be rejected, it is likely that a perturbing mechanism is responsible for
the extreme kinematics.  The apparent connection between the radio
and the kinematic properties  suggests that such a
mechanism could be shocks generated by the interac\-tion between the radio
outflow and the gas in situ
 (van Ojik et al.
1997). 
Galactic winds (Heckman, Armus \& Miley 1990) are an alternative
possibility.

In addition to these regions,
low surface  brightness Ly$\alpha$  halos  extending 
 beyond 
the radio structures have been detected in some HzRG. The kinematic
properties of such halos have been studied in detail only in one case, the
radio galaxy MRC1243+036 (van Ojik et al. 1996). The halo shows quies\-cent kinematics 
[FWHM(Ly$\alpha$)$\sim$ 250 km s$^{-1}$]   
compared to the  regions inside the radio structures [FWHM(Ly$\alpha$)$\sim$1200 km
s$^{-1}$]. The authors propose that the halo is a large rota\-ting gaseous disc
originating from the accretion associated with the formation of the galaxy.

Such quiescent low surface brightness halos (LSBHs, hereafter) 
are important since they show the gas properties unaffected by  kinematic perturbations.
 We are undertaking a research program using high S/N Keck spectroscopy
of a sample of high redshift radio galaxies ($z\geq$2.5)
whose goal is to search for 
kinematically unperturbed Ly$\alpha$ halos in HzRG. We will study the kinematic
and  ionization properties, as well as observed properties such as surface brightness,
size and luminosity. Constraints  will also be set 
 on the halo
mass  and its possible origin.

 The results on  the radio galaxy 
USS0828+193 are presented in this paper. In a forthcoming paper we will discuss
the results on the rest of the sample.

\section{Observations and data analysis}

USS0828+193 (R\"ottgering, Miley \& Chambers 1995) ($z=$2.57) is  a large radio source 
($\sim$70 kpc)\footnote{We assume $H_0$=72 km s$^{-1}$ Mpc$^{-1}$ and $q_0$=0.5. In this
cosmo\-logy, 1 arcsec corresponds
to $\sim$5 kpc} showing
a double morphology and a jet extending from the core to the northern hot spot.
The optical HST image   shows 
an irregular morphology
consisting of several clumps aligned with the radio axis
 (Pentericci et al. 1999, see also Fig.~1 top panel in this paper). The high polarization
level ($\sim$ 10 per cent) implies that  the UV rest frame continuum 
 is dominated by continuum
from the active nucleus (AGN) scattered by (most probably) dust 
in the extended gas (Vernet et al.
2001).

	The spectra were obtained  with the Low Resolution Imaging Spectrometer 
(LRIS, Oke et al. 1995) with its polarimeter (Goodrich, Cohen, Putney  1995) at the Keck II 10 m telescope in December 1997. We
used a 300 line mm$^{-1}$ grating and 1 arcsec  wide slit which provide 
a dispersion of 2.4 \AA ~  pixel$^{-1}$ 
and an effective resolution of
  FWHM$\sim$10.5 \AA  ~ (instrumental profile).  The exposure time was 5 hours.
The seeing  ($\sim$1.0 arcsec) was measured with the M5V type star located inside the slit.
 The slit was  oriented along the radio axis, with PA 44$^{\circ}$ (Carilli et al. 1997).
For a more detailed description of the observations  and data reduction 
see Vernet et al. 2001.

 Each individual frame was calibrated in
wavelength and corrected for slit curvature using arc spectra. This
initial wavelength calibration was refined using strong sky lines. In
order to compensate for residual  distortions (shifts) due to flexures in
LRIS, we extracted from every frame a section around Lya+NV 1240 and one
around CIV 1550+HeII 1640 (CIV and HeII respectively, hereafter). The individual sub-frames corresponding to the
same  spectral range were spatially 
aligned (using the continuum centroid) and combined. The two final frames (one for
Ly$\alpha$+NV and another one for CIV+HeII) were also spatially aligned, assuming
that the continuum centroid has the same spatial position at all wavelengths. This is
reasonable in this small spectral range, taking also into account that the spatial
resolution element has a large physical size ($\sim$ 5 kpc).

	The  spectra (Ly$\alpha$+NV and HeII+CIV frames) 
were divided in several apertures along the spatial
direction and a spectrum was extracted for each one. The apertures were 
selected so that the gas   has apparently similar 
kinematic properties across the spatial extension  of a given 
aperture. The apertures were also selected  to obtain enough S/N ratio in Ly$\alpha$, CIV and HeII
 (if possible) 
to fit
the line profiles (see Figs.~3 and 4). Separate apertures were  defined for the gas beyond the
radio structures.

\section{Results}

\subsection{Radio/Optical overlay}

	Figure 1 (top panel) shows
the overlay between  the optical WFPC2 HST image  and  VLA contours
(see Pentericci et al. 1999 for a description). The
 2-dimensional Ly$\alpha$
Keck 
spectrum is shown on the bottom panel spatially aligned with the image.
We identified the radio core with the 
brightest component in the NICMOS HST  image. Our main conclusions would not be affected
if we chose to identify the radio core with the vertex of the ionization cone
described by Pentericci et al. (1999) or the brightest optical component.

 An interesting feature is already obvious in this figure:
  a low surface brightness Ly$\alpha$ halo extending 
for $\sim$ 16 arcsec or  80 kpc inside and beyond the radio structures. It extends to
much larger distances than the material observed on the HST images ($\sim$4 arcsec).
The HST image shows, therefore,  only a small fraction of the total
 distribution of material. The kinematics of this halo is more quiescent than the
kinematics of the brightest regions.

\begin{figure}
\includegraphics{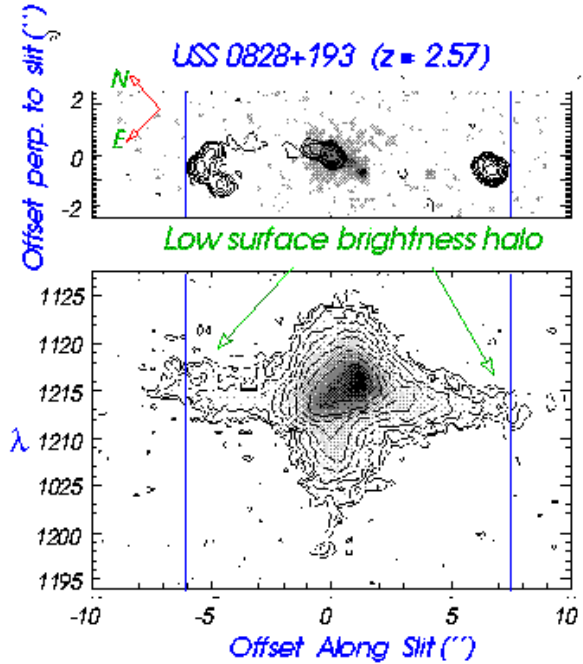}
\vspace{3.7in}
\caption{Top:  overlay between  the optical HST image and 
the radio VLA contours. The 2-dimensional spectrum of the Ly$\alpha$ emission line 
(bottom panel) is spatially aligned with the radio/optical images.
The spatial zero is the position of the continuum centroid measured on
the Keck spectra. The  vertical lines indicate the outer edges of the
radio lobe.}
\end{figure}

\begin{figure}
\includegraphics{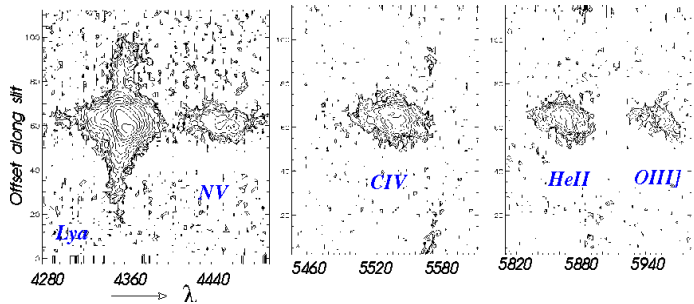}
\vspace{2in}
\caption{Contour plots of the main emission lines.}
\end{figure}

\begin{figure}
\includegraphics{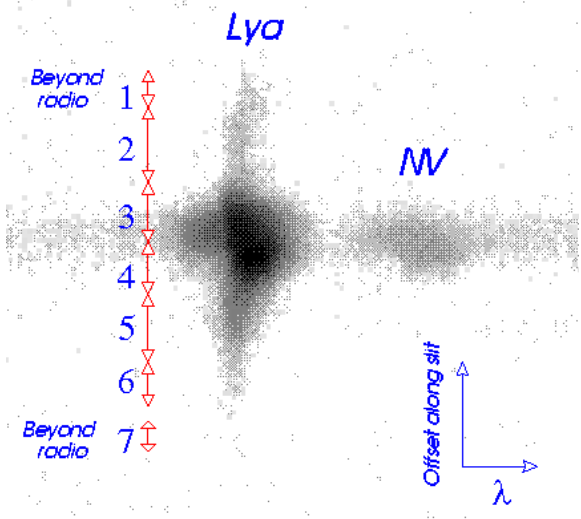}
\vspace{3in}
\caption{2-dimensional Ly$\alpha$+NV spectrum and selected apertures.}
\end{figure}

\subsection{The gas kinematics}

	We show in Fig.~2 the 2-dimensional spectrum (contours) of the main emission lines. 
Seven apertures were selected in the spatial direction for the kinematic analysis, which are indicated in 
Fig.~3. The corresponding 1-dimensional spectra in the Ly$\alpha$  and the  CIV+HeII regions are
shown in Fig.~4.

	Apertures $1, 2, 6, 7$  correspond to the low surface brightness 
 Ly$\alpha$ 
halo. Although very noisy, CIV and HeII are detected in 
apertures $2$ and $6$ (see Fig. 4).

 The Ly$\alpha$ and HeII spectral profiles were analysed and fitted in each aperture
with one, two or three Gaussians, depending on the quality of the fit.  CIV, in spite of being a strong line, 
turned out to be very difficult to constrain due to the
presence of the doublet components  (for a discussion on the uncertainties on 
interpreting the kinematics
of the extended gas in HzRG using rest frame UV emission lines see Villar-Mart\'\i n
et al. 2001). Therefore the results of the multiple component fit
are only shown for Lya and HeII. Those presented for CIV correspond to a
single Gaussian fit. For each kinematic component,
the FWHM (corrected for instrumental broadening in quadrature) and the velocity shift relative 
to the emission
at the continuum centroid were calculated (the HeII line
at  the continuum centroid was used to estimate the zero velocity). Some of the fits are shown in Fig.~5.

\begin{figure*}
\includegraphics{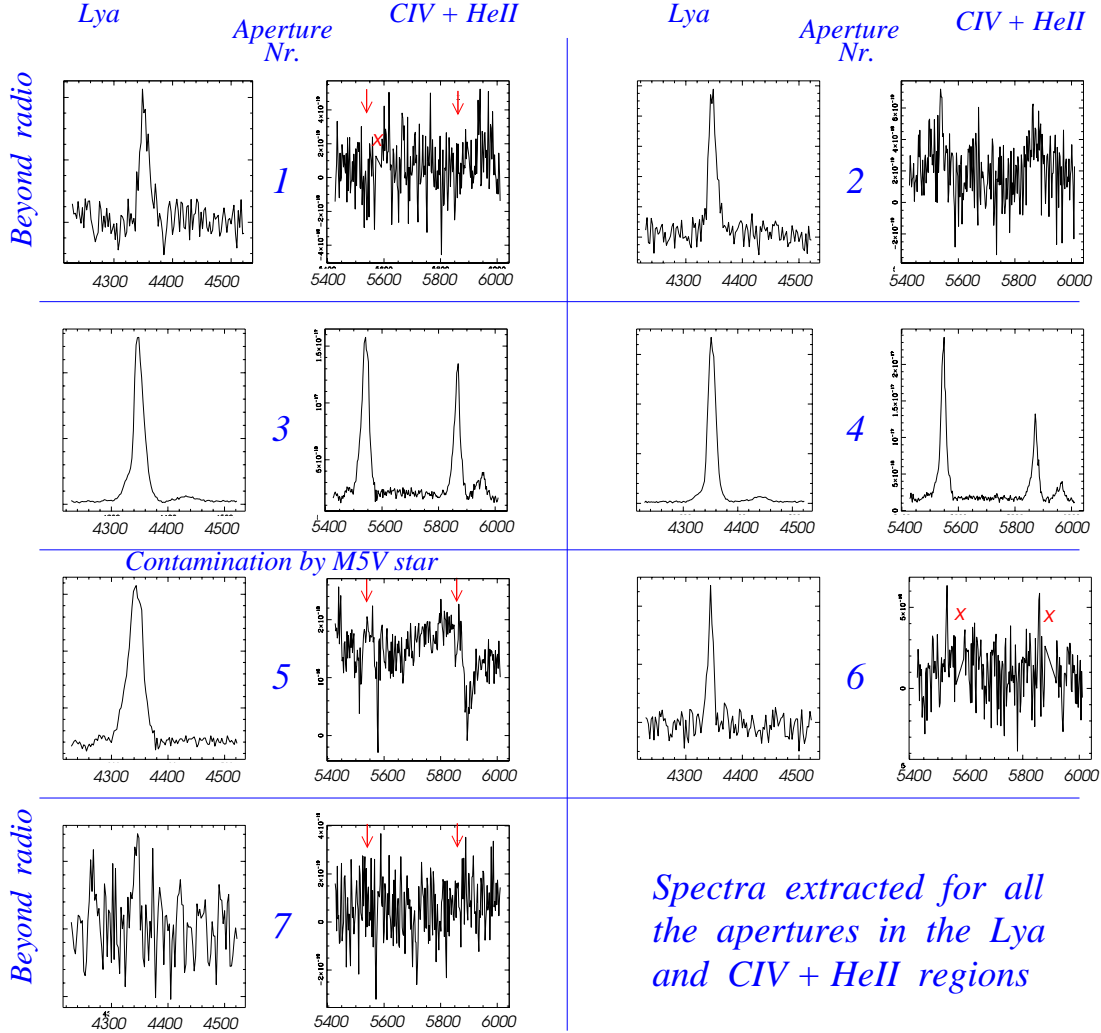}
 \vspace{6in}
\caption{Spectra extracted for the seven apertures indicated in Fig.~3. The expected 
position for the
CIV and HeII lines has been indicated with open arrows in the noisiest spectra.
CIV and HeII are detected in the low surface brightness halo (see spectra of
apertures $2$ and $6$).
Therefore, the gas is ionized and contains heavy elements. 
There is marginal evidence for continuum
detection in the halo. The interpolation (indicated with 'x') on the red
side of CIV and HeII in apertures  $1$ and $6$ was done to remove  
residuals from sky emission lines.}
\end{figure*}

\begin{figure*}
\includegraphics{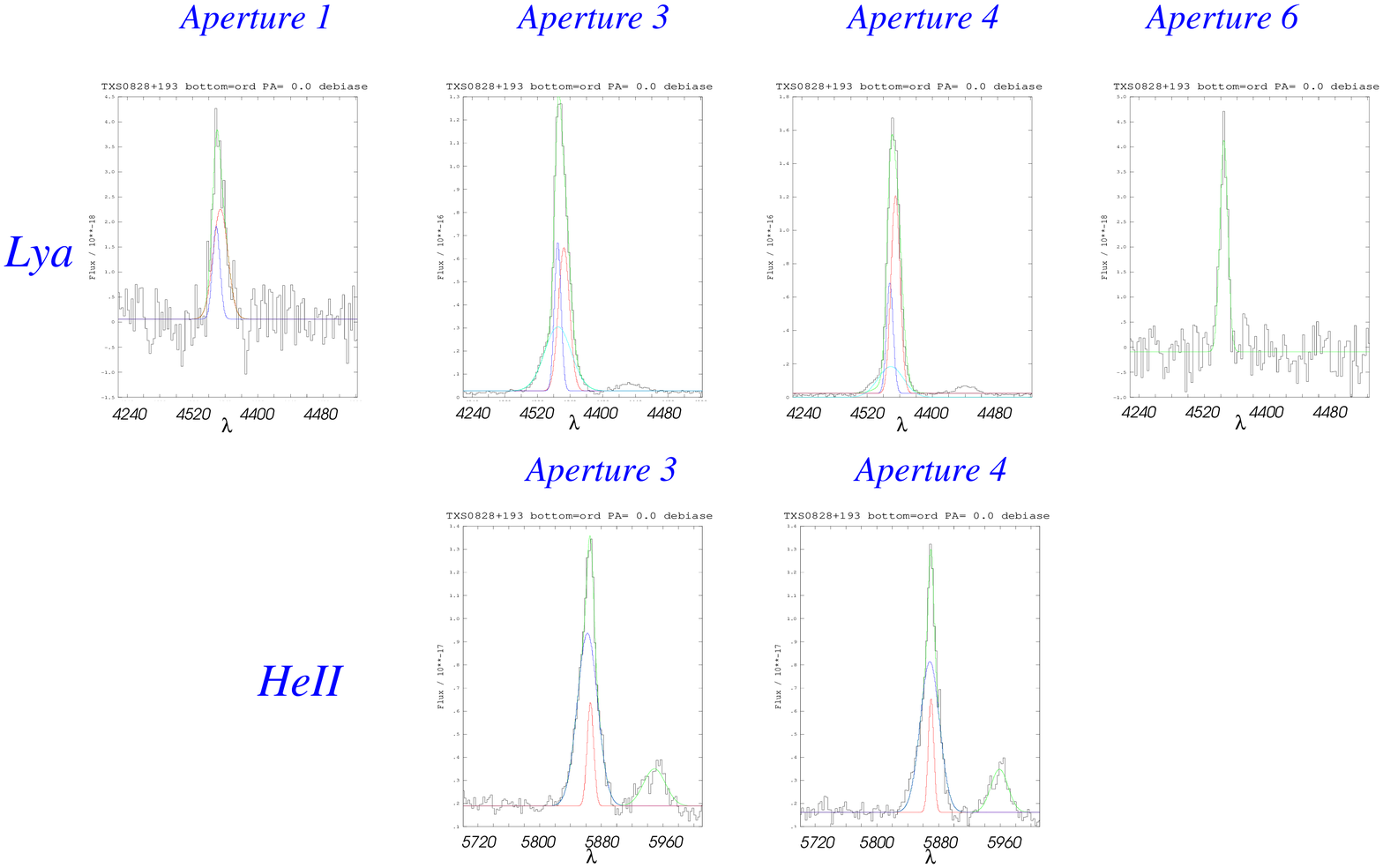}
\vspace{4in}
\caption{Examples of some of the fits to the Ly$\alpha$ and HeII lines. The fit and the individual
components are shown.}
\end{figure*}

Three kinematic components are found in Lya: (i) a narrow component with
FWHM$<$400  km s$^{-1}$  detected in all apertures; (ii) a broad component with
FWHM$\sim$1200  km s$^{-1}$  detected in all apertures except $6$ and $7$ (note this is
also detected beyond the radio lobe in aperture $1$); (iii) a very
broad component with FWHM$\sim$2500 km s$^{-1}$ only found in apertures $3$ and $4$
(brightest regions).  The best fit to the HeII profile  
reveals two components in apertures $3$ and $4$ similar (in FWHM) to components (i) and (ii)
 in Ly$\alpha$. Narrow HeII (aperture $2$) and CIV (apertures $2$ and $6$) 
 are also detected.

{\it (i) The narrow component ($<$400 km s$^{-1}$):} 
 The spatial variation of the kinematic properties of this
component are shown on panels $a$ (FWHM)  and $b$ ($V_S$) in Fig.~6. 
Diffe\-rent symbols
 are used for Ly$\alpha$ (circles), HeII (stars) and CIV (triangles).   This component  
 extends  across the whole object ($\sim$80 kpc) 
and beyond the radio structures. 
The kinematic properties are very uniform across the whole spatial extension.
There are no obvious changes associated with the radio structures.

Ly$\alpha$ is very
sensitive to absorption by neutral hydrogen and the interpretation 
of the spectral profile in terms
of pure kinematics is risky. In fact, van Ojik et al. (1997) found that 
the entire blue wing of the
Ly$\alpha$ (spatially integrated) profile of this galaxy is absorbed by neutral gas 
associated with the
galaxy. However, the fact that a similar narrow component has been found in HeII (which is not
a resonance line) strongly supports
the existence of a narrow component in Ly$\alpha$ in addition to the broad components.  
This is also supported
by the  very narrow profile of CIV and HeII  in apertures $2$ and $6$  and
the very narrow peak observed on top of the CIV and HeII lines in the inner apertures,
 $3$ and $4$ (see Fig. 4).

\begin{figure}
\includegraphics{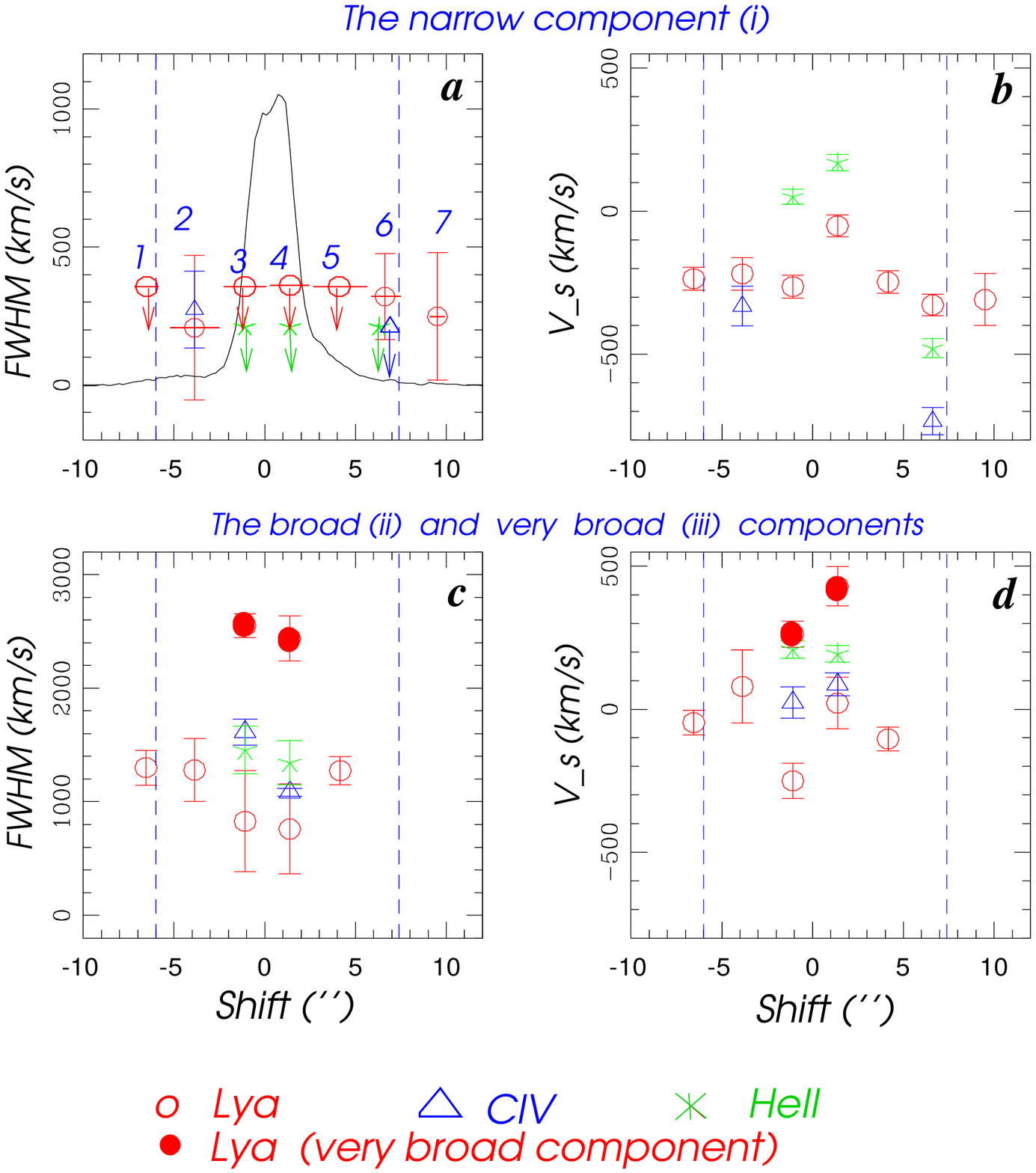}
\vspace{4in}
\caption{Kinematic properties of the individual components detected in
the different apertures. The FWHM appears on the left panels and
the velocity shift   relative 
to the HeII emission
at the continuum centroid is  plotted on the right panels. 
The narrow component is plotted in panels $a$ and $b$ and  the broad 
and very broad components are plotted in 
panels $c$ and $d$. 
 The dashed vertical lines mark the outer edge
of the radio lobes. Arrows indicate upper limits. The horizontal  lines in panel $a$
show the spatial extension of the aperture indicated with a number.
 Notice the
uniform properties of the narrow component  across the whole
object and the similarity with  low redshift radio galaxies kinematic properties. 
The detection
of the broad component (panels $c$ and $d$) beyond the radio structures is 
also interesting,
since it suggests that jet-cloud interactions are not responsible for the perturbed
motions.}
\end{figure}

	{\it (ii) The broad component   ($\sim$1200 km s$^{-1}$)}  
(open circles for Ly$\alpha$, open triangles for CIV and stars for HeII 
in pa\-nels $c$ and $d$ in Fig.~6). 
This component is also spatially extended  ($\sim$11 arcsec) and  with quite
uniform FWHM across the object, with no apparent link with the radio structures.  
This kinematically perturbed gas
is also found in the Ly$\alpha$ halo. The  measured 
FWHM is consistent within the errors for the three emission lines (HeII, CIV, Ly$\alpha$) where it is
detected, with marginal evidence 
for  narrower Ly$\alpha$. This could be due to absorption by neutral
hydrogen. This component extends beyond the radio 
structures (aperture $1$). Seeing effects 
might be responsible for spreading the detection of this component beyond the radio lobes.
Astrometry errors are not likely to be responsible (see below).

	{\it (iii) The very broad component   ($\sim$2500 km s$^{-1}$)}
(solid circles in panels $c$ and $d$
in Fig.~6)  is spatially extended  and it is only detected in Ly$\alpha$ in apertures $3$ and $4$. 

\section{Discussion}

\subsection{The nature of the narrow component}

This component (which extends across the entire object and beyond the edge of
 the radio lobes) shows  
  no apparent association with the radio structures.  
The kinematic properties  are similar to those of low redshift radio galaxies
(FWHM and velocity shift of a few hundred km s$^{-1}$, e.g. 
Tadhunter, Fosbury \& Quinn 1989).
This
 suggests that there is no (or little) 
interaction and this gas is unperturbed
by the jet/radio lobes.  Absorption of Ly$\alpha$ photons could explain the difference in the velocity
shift of Ly$\alpha$ and HeII in the inner apertures. 
 The kinematics of this narrow component is consistent with gravitational motions.

	This gaseous component also emits  CIV and HeII lines. It is the first time that 
lines other than Ly$\alpha$  are detected from the
low surface brightness halo of a HzRG.   This implies that the gas is ionized.
It is not a 'mirror' of neutral hydrogen
 that reflects Ly$\alpha$ by resonant scattering 
(Villar-Mart\'\i n, Binette \& Fosbury 1996).

There is marginal evidence for continuum detection in the halo (see Fig.~4).
The nature of this continuum is very uncertain. It could be due to scattered light, although
stars cannot be rejected.
The large equivalent widths of CIV and HeII  suggest that the gas
is not ionized by stars, but rather by a mechanism related to the active nucleus. 
This is
also implied by the 
CIV/HeII$\sim$1 value  measured for the narrow component  (at
least in aperture $6$ and maybe also aperture $2$, see Fig~4), similar to the 
ratio measured in 
aperture $3$ and consistent with measurements in
other HzRG (e.g. Villar-Mart\'\i n, Tadhunter \& Clark 1997).
 This   suggests the same ionization mechanism in the low surface brightness halo and
the bright regions, in spite of the different
kinematic proper\-ties.  Since the halo extends beyond the
radio structures where shocks are unlikely to be present, the ionization
mechanism is probably ionization by the central AGN (ionization by the 
continuum generated by hot shocked gas is an alternative mechanism). 

On the other hand, the detection of
CIV emission shows that the halo has been enriched with heavy elements at very large distances
from the nuclear region. It is not possible to constrain the metallicity using only
the CIV/HeII  ratio, but solar abundances are certainly possible (Villar-Mart\'\i n, Tadhunter \& Clark   
1997, Vernet et al. 2001). A galactic wind, whose presence is suggested by
the broad component detected at similar distance from
the nucleus, might be the mechanism responsible for the enrichment of the gas.

\subsection{The nature of the broad (ii) and very broad (iii) components}

	The broad component ($\sim$1200 km s$^{-1}$) is apparently detected beyond
the Eastern radio lobe. Errors in the Astro\-metry are unlikely to be responsible:
if the active nucleus were located at the vertex of the ionization cone described by
Pentericci et al. (1999) or at the brightest optical feature,  the broad component
would extend even  further beyond the edge of the radio lobe. However, we do not discard the possi\-bility
that seeing effects might be smearing the emission enough that the broad component actually does not
extend beyond the radio lobes.

	If the detection beyond the radio structures is real, 
this   implies that  the motions of this gas are not a consequence of jet-cloud interactions. 
Evidence for perturbed gas beyond radio structures 
has been  found in other objects (e.g. Villar-Mart\'\i n, Binette and Fosbury 1999, 
Sol\'orzano-I\~narrea, Tadhunter \& Axon  2001).
Galactic winds could be an alternative possibility.  Clouds exposed to ram pressure of the wind 
 can be accelerated to  several hundred km s$^{-1}$ and shifted by $\sim$1000 km s$^{-1}$
(Heckman, Armus \& Miley 1990).
The integration along the line of sight  through several  such clouds could explain the broad FWHM of
this component. Gas free-falling into the galaxy during the formation process from a large distance 
can also explain FWHM of this order (Lehnert \& Becker 1998). 
The nature of the very  broad component ($\sim$2500 km s$^{-1}$)
 in apertures $3$ and $4$ is not clear. It might be scattered 
light from a hidden broad line region
(Vernet et al. 2001).

\subsection{Mass and density of the ionized gas in USS0828+193}

(Hereafter, we will refer to the gas emitting the  narrow component
as the low surface brightness halo, LSBH).
 
The integrated Ly$\alpha$ flux $F_{Ly\alpha}$ of the LSBH along the slit
is 1.64$\times$10$^{-15}$ erg s$^{-1}$ cm$^{-2}$, which corresponds to
a luminosity $L_{Ly\alpha}$=3.8$\times$10$^{43}$ erg s$^{-1}$. For 
 pure case B recombination and $T=$10$^4$ K, $L_{Ly\alpha}$ is given by
 (McCarthy et al. 1990a):

\begin{equation}
L_{Ly\alpha} = 4 \times 10^{-24} n_e^2~f~V ~{\rmn erg ~s^{-1}}
\end{equation}

\vspace{0.3cm}

where $f$ is the volume filling factor of the gas, $n_e$ the electronic
density and $V$ is the total volume inside the slit.

	We have calculated the volume assuming that the gas of the LSBH is
ionized by the active nucleus (see \S 4.1) and is homogeneously distributed.  
In this scenario, the ionized halo has a biconical
geometry. We have assumed that the cone axis is on the plane of the sky with an opening angle 
for the individual cones of $\sim$90$^{\circ}$
(Barthel 1989). The 2-dimensional spectra show  an extension of $\sim$8 arcsec (40 kpc)
 along the slit from the AGN. 
The volume of the bi-cone slice included in the slit (1 arcsec
wide or 5 kpc) is $\sim$3.2$\times$10$^{68}$ cm$^3$. 
Assu\-ming $f=$10$^{-5}$ (McCarthy et al. 1990a), we obtain $n_e \sim$50 cm$^{-3}$. 
This is a very crude estimate. Large uncertainties are due to: a) 
 possible scatter on the    $f$ values.
Values for  $f$ of the order of 10$^{-7}$-10$^{-8}$ have been estimated for Ly$\alpha$
nebulosities associated with high redshift quasars (Heckman et al. 1991) and some low
redshift radio galaxies (Clark et al. 1998); b) 
the fact that the measured Ly$\alpha$ luminosity is probably  lower
than the case B value (due to the dust and/or neutral hydrogen absorption) and c)
the assumptions on the geometry.
The estimated $n_e$ value is likely to be a lower limit for the density and it could
be at least one order of magnitude higher.

The mass of ionized gas in the LSBH is given by

\begin{equation}
M(H^+) = n_e \times V \times m_p \times f
\end{equation}

where V is in this case the total volume of ionized gas and $m_p$ the mass
of the proton. (We assume that the density is constant across the halo for
simplicity, but it is possi\-ble that the density increases inwards). Considering the
two cones we obtain a  total volume of 2.3$\times$10$^{69}$ cm$^3$  and
therefore $M(H^+) \sim$9.6$\times$10$^8$ $M_{\odot}$, for $n_e$=50 cm$^{-3}$. 
This is proba\-bly un upper limit. Since $M(H^+) \propto f^{1/2}$ and taking into account the
uncertainties on $f$, the mass could be at least ten times lower.

The halo material outside the 
ionization cones is likely to be neutral (or have a low ionization level due to, for ins\-tance, metagalactic radiation). 
Evidence for reservoirs of neutral gas 
associated with distant radio galaxies has been found by van Ojik et al. (1997).
The authors  found that the Ly$\alpha$ profile in  eleven (including USS0828+193) 
out of eighteen distant radio galaxies
shows deep troughs which they interpret as HI absorption by absorbers that 
are  likely to be physically associated with the galaxy hosting 
the radio source or its direct
environment \footnote{ An interesting case is the radio galaxy 
MRC0943-242 ($z=$2.92), which 
shows large scale absorption troughs (R\"ottgering et al. 1995, Binette et al. 2000), both
in Ly$\alpha$ and CIV. Binette et al. concluded that the  absorbing gas is metal poor ($Z \sim$ 0.01 $Z_{\odot}$)
and is located much further  outside the zone of influence of the radio jet cocoon}.

	The LSBH in USS0828+193 (and other HzRG) could be part 
of a gaseous reservoir that surrounds the object completely. The gas outside the ionization cones 
 is neutral and we see it in absorption   and  the gas inside the  cones is 
ioni\-zed and  we see it in emission.
If we assume that the neutral gas of the halo has the same density and 
filling factor as the ionized fraction, we obtain  
$M(H^0) \sim$2.4$\times$10$^9$ $M_{\odot}$.
The mass of ionized ($T\sim$10$^4$ K) and neutral gas in 
the halo is therefore $M_{H^++H^0} \sim$3.4$\times$10$^9$ $M_{\odot}$.

	Following a similar procedure, we obtain $n_e \sim$ 150 cm$^{-3}$ for the
broad line emitting gas (the same $f$ and cone opening
angle were used). To calculate
the volume, two cones of height 4 and 6 arcsec
respectively were considered. These are the projected distance values at 
which the broad component is
detected at both sides of the continuum centroid (see Fig.~6). This gives a volume
of $\sim$1.3$\times$10$^{68}$ cm$^{3}$ inside the slit and 1.2$\times$10$^{69}$ cm$^{3}$
as the total volume inside the cones.
The  mass of broad emission line gas inside the ionization cones is then 
$\sim$1.5$\times$10$^9$ M$_{\odot}$.

	 We conclude that a large fraction of the gaseous  (with $T \leq$10$^4$ K) 
mass in HzRG  might be  
in large halos that su\-rround the entire object.  Part of the  halo (the 
gas inside
the ionization cones) is ionized
and we see it as low surface brightness line emission.  The rest 
  is neutral (or with very low ionization level) 
and we see it as absorption features imprinted on the Ly$\alpha$  profile

\subsection{The origin of the LSBH in USS0828+193}

We have considered several scenarios:

\begin{itemize}

\item A rotating disc (evidence for large discs associated with low redshift
radio galaxies is discussed in \S4.5.1).
Signs for rotation of the extended gas in nearby radio galaxies
have been found by several authors (Heckman et al. 1985, Baum, Heckman \& van Breugel 1990, Tadhunter,
Fosbury \& Quinn 1989).  In this case, $M_{rot} =  \frac{R~V^2}{G~sin~i^2}$
where $R$ is the radius of the disc (40 kpc), $V$ is half the amplitude of the rotation curve and $i$ is the inclination
angle of the disc with respect to the plane of the sky. This formula is applicable
when we measure the rotation along the line of nodes. 
Velocity curves that resemble
rotation have been found in several HzRG, but it is not clear whether this is an
effect of multiple kinematic components (Villar-Mart\'\i n et al. 2001).

	The current data do not allow us to set any constraint on the mass if the
halo is a rotating disc:  we do not have information on
 the inclination angle of the disc (imaging would help
to constrain it), neither on the direction of the line of nodes, which 
 is likely to 
be misaligned with respect to the slit position. 
Van Ojik et al. (1996) propose that the giant Ly$\alpha$ halo associated with
the radio galaxy MRC 1243+046 ($z=$3.6)  has settled as a  rotating disc
whose orientation (the plane of the disc) is within $\sim$20$^{\circ}$ of the orientation of the radio axis.
However, this is contrary to the 
 trend observed in low redshift powerful radio galaxies
with signs of rotation. In such objects,  
the rotation axis of the gas and the radio axis 
trend to be aligned within 
20$^{\circ}$ (Heckman et al. 1985).
If this is
the case for the LSBH in USS0828+193, the rotation axis would be 
approximately  along  the slit, since this was aligned with the radio axis.

	The Ly$\alpha$ velocity curve of the narrow component in USS0828+193 does 
not resemble 
a standard Keplerian  rotation curve for a slit located along the line of nodes.
However, it is interesting to note that  the observed pattern is similar to what we
expect from a rotating disc
 if the slit was not aligned  with the line of nodes, but rather close to the 
perpendicular direction and with some impact 
parameter with respect to the rotation center 
(i.e. there is a distance between the center of the slit and the center of rotation). 
Such a pattern is expected if the disc has a  certain inclination relative to the plane
of the sky.
A close analysis
of the CIV line profile in the two inner apertures ($3$ and $4$, see Fig.~4) reveals 
the presence of a very narrow
peak (on top of a domi\-nant broader component). The velocity shift of this narrow
peak has been plotted in Fig.~ 7 (solid triangles); CIV  shows a 
 pattern similar to Ly$\alpha$.

3-dimensional spectroscopy
would be very valuable to obtain information about the geometry  and the kinematic pattern along different directions across the halo.

\begin{figure}
\includegraphics{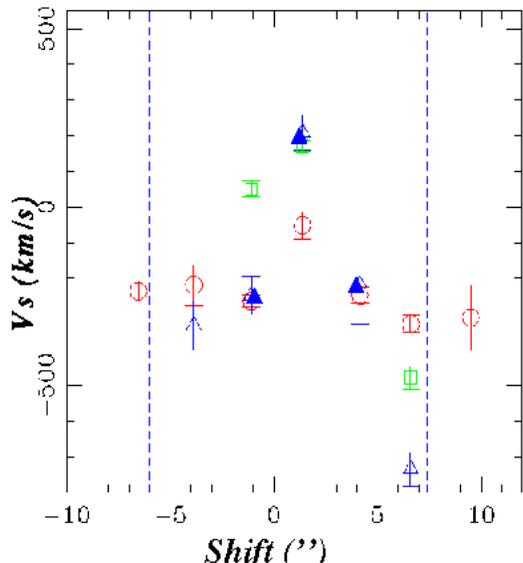}
\vspace{3in}
\caption{Same as panel $b$ in Fig.~6  but with the measured velocity shift for the
very narrow peak observed in CIV in the inner apertures (solid triangles). CIV follows
a similar pattern as Ly$\alpha$.}
\end{figure}

\vspace{0.5cm}

\item A virialized system: The halo could be a virialized system 
that consists of $N$ individual clouds  of mass $m$ that
move with  random motions \footnote{Pettini et al. (2001) also calculate 
dynamical masses of Ly break galaxies  in terms of gas clouds.}. 
This could represent the scenario
where the halo is formed by galactic satellites.
Applying the virial theorem $M_{dyn} = \frac{5~R~V_v^2}{G}$ (Carroll \& Ostlie 1999),
 where $R$ is the radius of the system (40 kpc)
and $V_v$ is the radial velocity  dispersion of the clouds within the halo (given by the $\sigma$ of the emission
lines $\sim$130 km s$^{-1}$) we obtain $M_{dyn}$ = 8$\times$10$^{11}$ M$_{\odot}$ 
(this is the total mass of the system, luminous+dark).

\item Inflow: In this case $M_{inf} =  \frac{R~V_{R}^2}{2~G}$, where $V_R$ is the  infalling velocity 
of a particle at distance $R$ from the central mass $M_{inf}$. We have assumed 
 that the axis of the ioni\-zed cones (opening angle $\theta$=90$^{\circ}$)
 is on the plane of the sky.
The emission at each spatial position along the slit results from the integration of the gas emission within the
cone slice inside the
slit and 
along the line of sight (see Fig.~8). 
In such case, the FWZI of the narrow 
emission line at a given slit position gives the maximum  projected velocity displacement between the gas at 
the two opposite edges of the slice cone in that spatial position, i.e.,
FWZI = $V_{02} - V_{01}$ = 2$\times~V_{R}~sin(\theta/2)$, where $V_{02} = -V_{01} = V_R ~sin(\theta/2)$.
On the other hand, $R=\frac{r}{cos(\theta/2)}$, where $r$ is the observed projected 
radius. 
Therefore  

\begin{equation}
M_{inf} =  \frac{r}{cos(\theta/2)} ~ (\frac{FWZI}{2~sin(\theta/2)})^2  ~ \frac{1}{2~G}
\end{equation}
 In USS0828+193, FWZI$\sim$430 km s$^{-1}$ at $r=$40 kpc (8 arcsec)
as measured from the spectrum.
We obtain $M_{inf} \sim$ 6$\times$10$^{11}$ M$_{\odot}$  for the total mass of the system.

Values of this order  
have been estimated for the masses of
 low redshift radio galaxies and are  smaller than typical masses of  cD elliptical galaxies at
low redshift (Tadhunter, Fosbury \& Quinn 1989). If this result is confirmed, we would conclude that 
USS0828+193 will not become like the  
cD galaxies we see at the present epoch,  unless mergers are involved. 

\begin{figure}
\includegraphics{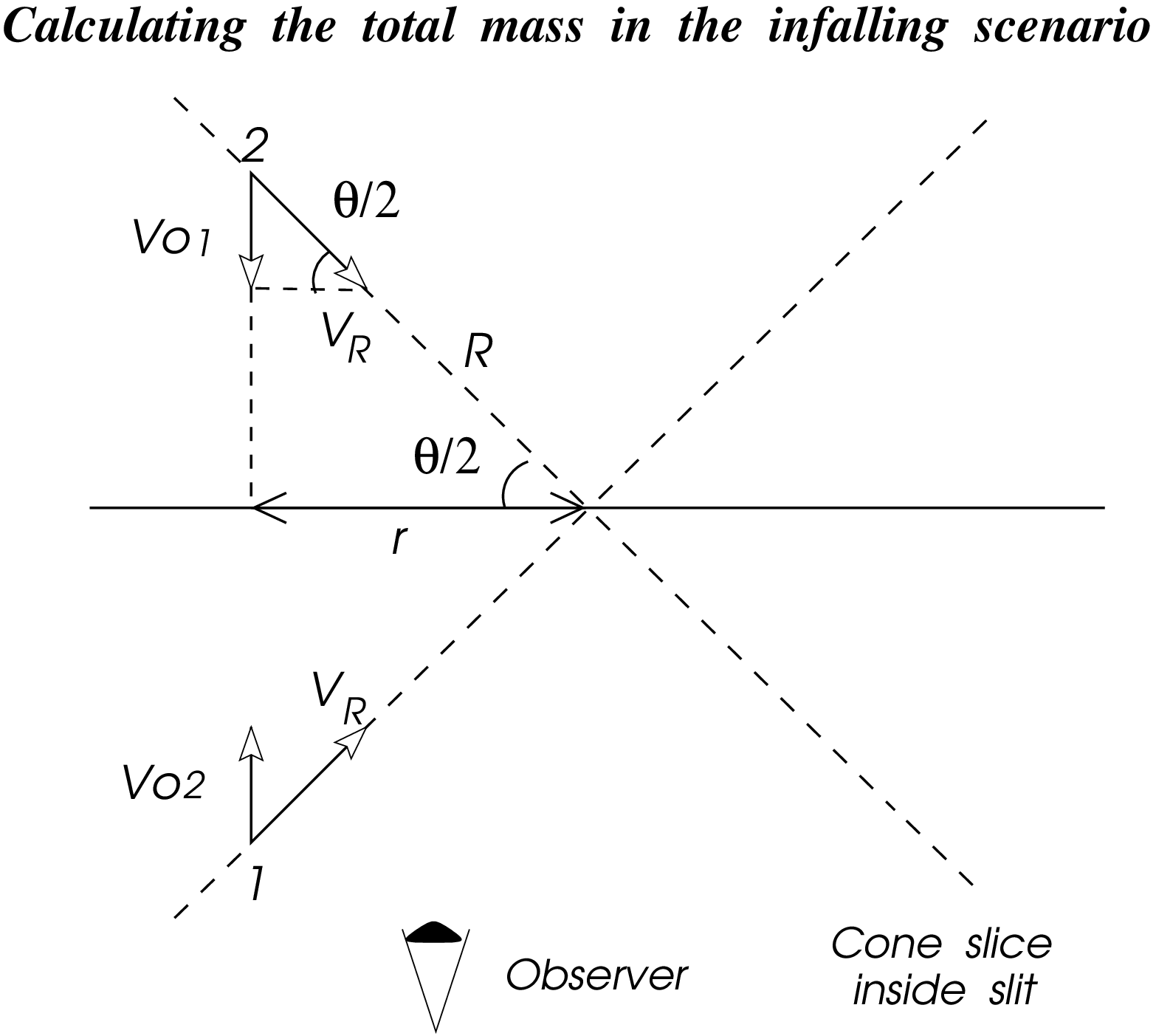}
\vspace{3.5in}
\caption{Infall scenario: $Vo_1$ and $Vo_2$ are the velocities of particles in positions 1 and 2, projected on the line
of sight (l.o.s) and $V_R$ is the radial (infalling) velocity towards the center.}
\end{figure}

\vspace{0.3cm}

Is it possible that the material in the halo is falling towards the center in a cooling flow manner 
(rather than in free fall) ? \footnote{In the free fall scenario, the only force acting on the gas is gravity.
In the cooling flow scenario, the pressure gradient in a hot corona acts against gravity}.
We have investigated whether  this scenario is plausible. We have assumed  that the emission line gas is
in pressure equilibrium with the hot phase. Typical temperatures of $T_{clouds}=$10 000 K and $T_{hot}=$10$^7$
for the optical emitting gas and the hot phase respectively and $n_{clouds}$=50 cm$^{-3}$ for the
emission line gas (\S 4.2), give $n_{hot}$=0.05 cm$^{-3}$ (the  uncertainties on $n_{clouds}$
and  $T_{hot}$ allow only crude estimates in the discussion that follows).

Adopting a value for the cooling function $\gamma$(T=10$^7$ K) = 5$\times$10$^{-23}$ erg
cm$^3$ s$^{-1}$ (Raymond, Cox \& Smith 1976), and combining the three equations in \S 3 (Nulsen, Stuart
\& Fabian 1984) the  mass of hot gas in the halo (inside the visible radius of 40 kpc)
 of USS0828+193 is $\sim$4$\times$10$^8$ M$_{\odot}$
and its  radiative cooling time
is $\sim$6$\times$10$^7$ years. This cooling time is rather short (much shorter than
the age of the Universe at that time). 	 On this basis, it is possible that 
 the gas we observe from the LSBH has cooled from the hot phase. 

	The mass that has cooled down is $\sim$10$^9$ M$_{\odot}$ (considering only the
ionized gas in the LSBH [see \S 4.2]). 
If this happened in $\sim$6$\times$10$^7$ yr,
the mass deposition rate is 17 M$_{\odot}$ yr$^{-1}$ (va\-lues of this order have been measured in nearby 
cooling flow galaxies, Heckman, Armus \& Miley 1990)

\vspace{0.2cm}
	We have assumed that the density of the halo is constant. However, this is a very simplified scenario and
the density profile is likely to be centrally condensed   
(Haiman and Rees 2001, HR2001 hereafter). 
The cooling time varies
as $n_e^{-1}$ (Nulsen, Stewart \& Fabian 1984), therefore, the outer, less dense 
gas will need longer time to cool down.
As van Ojik et al. found, when we account for this effect, the cooling time is 
still rather short at large distances
(r=100 kpc) for a density behaviour of both $\propto r^{-1}$ and $\propto r^{-2}$  
(2$\times$10$^8$ and 2.5$\times$10$^8$ years
respectively).  There has been, therefore, enough time for the gas to cool down.

	In  a cooling flow scenario, we  expect very narrow emission lines at
large distances ($\leq$100 km s$^{-1}$ at $\geq$tens kpcs, Fabian et al.
1987))\footnote{Emission lines as
broad as several hundred km s$^{-1}$ are observed in low redshift cooling flow nebulae
(e.g. Fabian et al. 1987, Heckman et al. 1989). The authors propose that galaxy interactions may generate 
turbulence.}, a
chaotic pattern in the velocity 
curve and a trend of
the line width to decrease with increasing radius (Heckman et al. 1989). This behaviour 
is not obvious on the data but we are limited by 
the low spectral resolution.  The kinematic properties of the halo do 
not help to constrain the validity of the cooling
flow scenario. Higher resolution would help to measure the FWHM more accurately at different spatial position.

\vspace{0.2cm}
\item  Outflow:  the halo could have been deposited by massi\-ve outflows (winds could explain the
existence of heavy ele\-ments in the halo).   We have calculated some basic parameters
characterizing a superwind which could explain the kinematic properties of the LSBH in USS0828+193. 
The ener\-gy injection rate for
an energy-conserving bubble inflated by energy injected at a constant rate and expanding 
into a uniform medium
with an ambient density $n_0$ (cm$^{-3}$) is related to the radius (in kpc) and 
velocity (in 10$^2$ km
s$^{-1}$) by (Heckman, Armus \& Miley 1990)

\begin{equation}
dE/dt = 3 \times 10^{41}  r^2_{kpc} v^3_{100}  n_0~ { \rmn erg s^{-1}} 
\end{equation}

and the dynamical time scale of the nebula is given by

\vspace{0.2cm}

\begin{equation}
t_{dyn,6} \sim 10 ~r_{kpc}/v_{100}~ { \rmn Myr}
\end{equation}

	where $v_{100}$ would be halve the separation in velocity 
between the two components of the double peaked profile produced by the expansion.

	The resolution of the spectrum is too low to resolve the double peak that
the expansion of the bubble would produce. 
We have estimated that two unresolved
emission lines separated by a given velocity $\delta v$ would produce a line of 
FWHM $\geq$ 2$\times \delta v$ = 4 $\times v_{100}$ (all velocities in units of 100
km s$^{-1}$). 
Therefore, from the  measurements
in our spectrum ($\sigma\sim$130 km s$^{-1}$) we obtain $v_{100} \leq$0.76.

The estimated density $n_e =$50 cm$^{-3}$  is probably 
the density of the post-shock gas in the expanding bubble.
 $n_0$ in the $dE/dt$ equation is, however, the pre-shock density, which is likely to be much smaller 
 ($\sim$several cm$^{-3}$,  Heckman, Armus \& Miley 1990). For $n_e$=1 cm$^{-3}$ the implied values
of $t_{dyn}$ and $dE/dt$ are $\sim$530 Myr  
(the corresponding time scale for ARP 220 is $\sim$100 Myr,
Heckman, Armus \& Miley 1990) and 2.1$\sim$10$^{44}$ erg s$^{-1}$ res\-pectively. 
The required star forming rate is   SFR$\sim$500 M$_{\odot}$ and a supernova rate of $\sim$4 SN yr$^{-1}$. 
Typical SFR values for Ly break galaxies (dust corrected)
are several hundred  M$_{\odot}$ yr$^{-1}$ (e.g. Sawicky \& Yee 1998). 

The expected total infrared luminosity of such a system would be $L_{IR} \sim$3$\times$10$^{12}$ L$_{\odot}$. 
This would place USS0828+193 in the category of ultraluminous infrared galaxies.

The kinematic signature of an expanding
bubble (or biconical structure) would be a region of double peaked emission line
profiles. Higher spectral resolution might reveal this pattern.

\end{itemize}

\subsection{Is this a halo from which the galaxy is still forming ?}

	As Haiman \& Rees discuss (2001),  in the process of formation of a
galaxy, smooth gas collapses from the virial radius  to its final orbit
radius. Numerical simulations have revealed a more complex process, where
a fraction of the infalling gas forms smaller clumps early on; these clumps 
then progressi\-vely merge together, collide and dissipate to form larger 
systems. In these scenarios, a robust feature of galaxy formation is at 
least in the early stages a
spatially extended 
distribution of gas. HR2001 showed that if the halo 
is illuminated by  ionizing radiation from a central quasar, it should be
observed as an extended Ly$\alpha$ fuzz with surface brightness 
$\sim$10$^{-17}$ erg s$^{-1}$ cm$^{-2}$ arcsec$^{-2}$.

	We compare in this section the properties of the LSBH in USS0828+193 with
those expected for the extended gas condensation in the early stages of galaxy formation models.

\begin{itemize}

\item  {\bf Total  mass}: At $z=$3 (total) halo masses in the range $\sim$4$\times$10$^{10}$-10$^{13}$ M$_{\odot}$ are expected for  a virial temperature in the range 
 $T_{vir}$=2 $\times$ 10$^5$ K -- 2 $\times$ 10$^7$ K
(HR2001). 
We have obtained $M_{tot} \sim$6-8$\times$10$^{11}$ M$_{\odot}$ for
the two scenarios (infall and virialized system of clouds) where we have been able to
calculate the total mass of the system. This value lies in the
range predicted by the models.

\item {\bf F(Ly$\alpha$)$_{SB}$}: The expected surface brightness for the
 Ly$\alpha$ fuzz is in the range $\sim$10$^{-18}$-10$^{-16}$
erg s$^{-1}$ cm$^{-2}$ arcsec$^{-2}$ (HR2001). This is consistent with the measured
value in USS0828+193 $\sim$ 3$\times$10$^{-17}$ erg s$^{-1}$ cm$^{-2}$ arcsec$^{-2}$.

\item {\bf Linear size}:  The linear spatial extension of the Ly$\alpha$ emitting
gas would be a fraction of the virial radius $R_{vir} \sim$ 10 - 100 kpc (HR2001),
consistent with 
the linear scale of the halo (radius) in USS0828+193, $\sim$40 kpc.

\item {\bf Black hole mass}: 	According to HR2001 model, the minimum mass required to
ionize the halo is given by: 

\begin{equation} 
M_{bh} \sim 6 \times 10^8 M_{\odot}~ (\frac{M_{tot}}{10^{12} ~M_{\odot}})^{5/3} ~(\frac{1+z}{6})^4
\end{equation}

 Assuming  $M_{tot}$=8$\times$10$^{11}$ M$_{\odot}$ as above, we find a lower limit for the 
black hole mass of $\sim$5$\times$10$^7$M$_{\odot}$.  This value is somewhat lower 
than the black hole mass we
expect accor\-ding to
 Gu, Cao and Jiang (2001), who  found 
that the vast majority of the quasars in the 1Jy, S4 and S5 catalogues 
have typical black hole masses of $\geq$10$^8$ M$_{\odot}$. 

\item {\bf Heavy elements}:  One of our most important
results is the finding of heavy elements  in the LSBH. Any model of
protogalactic  halos should consider metal enrichment. This is mentioned
by HR2001, although not discussed in details.

\end{itemize}

	The properties of the low surface brightness halo in USS0828+193 
 are, therefore, consistent with the expected properties for the extended gaseous halo, a required
ingredient for galaxy formation models.

	However, the existence of the halo does not necessarily imply that USS0828+193 is a galaxy
in the process of formation. Similar
halos  have been found in lower redshift radio galaxies, where the galaxy is expected to be fully formed. 
An example is the radio galaxy 3C368 at
$z=$1 (Stockton, Ridgway \& Kellogg 1996). This object shows low surface brightness [OII] 3727 emission 
with apparently quiescent kine\-matics, in addition to the
bright kinematically perturbed structures. Other examples at lower redshift are 
3C34 ($z$=0.69), 3C435A ($z$=0.47) 
(see 2-dimensional [OII] 3727 spectra in  Sol\'orzano-I\~narrea, Tadhunter \& Axon (2001). The halos extend
for more than 100 kpc).
Such halos
might be similar to the LSBH we have discovered in the Ly$\alpha$ light in our more
distant radio galaxies. If this is the case, the existence of LSBH associated with distant radio galaxies
does not imply an early stage in the formation process. On the other hand, 
the consistency of the LSBH properties in USS0828+193 with expectations for the initial gas reservoir,
suggests that the halo gas is part of the material from which the galaxy
started to grow.

\subsubsection{Comparison with other objects}

	We discuss here several structures found around other types of galaxies at different redshifts which
might have the same origin as the low surface brightness halos found around distant
radio galaxies.

\begin{itemize}

\item Galactic envelopes: observations of  absorption line systems in the
spectra of background quasars have provided evidence for
large ($R\sim$100 $h^{-1}$ kpc) extended 
gaseous envelopes that surround galaxies
of a wide range of luminosity and morphological type (e.g. Chen, Lanzetta \& Webb 2001;
Lanzetta et al. 1995).  Chen, Lanzetta \& Webb (2001) have proposed  accreting
satellites as the origin of these structures  to explain the chemical enrichment at large distances.

\item Giant Ly$\alpha$ halos associated with Ly break galaxies: Steidel et al. (2000)
found two giant  (physical extent $\geq$100 $h^{-1}$) diffuse
 Ly$\alpha$ emitters apparently associated with previously known Ly break galaxies
at $<z>=$3.09. The origin of these structures is not clear. Proposed explanations are  a cooling flow nebula
 (Steidel et al. 2000) or 'hyperwinds' (Taniguchi, Shioya \& Kakazu, 2001).

\item Large scale ($\geq$several tens of kpcs) HI disc like  structures have been found in several low redshift radio 
(Morganti et al. 2002) and elliptical galaxies (Oosterloo et al. 2002). 
The authors propose that these structures are the results of mergers. The LSBH found around
 USS 0828+193 and other distant radio galaxies, might be the progenitors of the discs found at low redshift.

\end{itemize}

 Therefore, giant gaseous halos ($\geq$ several tens of kpc) are found to be often associated with  galaxies
in general: active and non active, of different morphological types and luminosities and at different epochs. The origin
and nature
of such halos is not clear and different explanations have been proposed for different objects, but 
the definitive answer is uncertain. 
An interesting possibility is that these halos have a similar origin related to the formation process of the
galaxies.

\section{Summary and Conclusions}

By means of the spectral decomposition of the emission line profiles in USS0828+193, 
we have
isolated the emission from a giant ($\sim$80 kpc) reservoir of apparently kinematically unperturbed gas, 
whose kinematics is consistent with being 
gravitational in nature. Emission lines other than Ly$\alpha$ (CIV, HeII) have been detected for
the first time in such a halo. This implies that the gas is ionized (probably by the continuum
from the AGN) and enriched with heavy elements. We find marginal evidence for continuum detection
from the halo. 

We propose that the low surface brightness halo  in USS0828+193 (and other HzRG) could be part 
of a gaseous reservoir that surrounds the entire object.
We have consi\-dered several possible scenarios to explain the origin of the halo. 
The gas could have settled in a rotating disc. This is suggested by  
the velocity pattern of the  halo emission, although 3-dimensional spectroscopy would be necessary to
set tighter constrains on the kinematic patterns.  Evidence for giant gaseous discs 
has been found by other authors in low redshift radio galaxies and elliptical galaxies.

	The halo could also be a group of individual clouds (galactic satellites?) in a virialized system. 
In this case,
the expected mass is $\sim$8$\times$10$^{11}$ M$_{\odot}$. A similar value is estimated if
the halo is infalling towards the center of the potential well. Such values are 
similar to estimates of masses of
 low redshift radio galaxies, and smaller than typical masses of  cD elliptical galaxies at
low redshift. If this result is confirmed, we conclude that USS0828+193 will not become like the
cD galaxies we see at the present epoch, unless mergers are involved.

	Alternative scenarios that we have discussed (also consistent with the observations)
are a cooling flow from a hot phase and outflows (galactic winds). In this last case, 
star forming rates $\sim$500 M$_{\odot}$ yr$^{-1}$ are required, consistent with  estimates for
Lyman break galaxies.

The properties of the low surface brightness Ly$\alpha$ halo detected in the radio galaxy
USS0828+193  are consistent  with the expectations for the original gaseous reservoir  from
 which the galaxy started to form. This does not necessarily imply that
the galaxy is in the process of formation.

	We have also discussed the possible link of the low surface brightness halos
 with a) the very extended envelopes found around
galaxies of  different  morphological types and luminosities
based on studies of absorption line systems in the spectra of background quasars 
b)  the giant Ly$\alpha$ halos associated with some Ly break
galaxies c) the very extended reservoirs (claimed to be discs) of neutral gas 
associated with some nearby radio galaxies and elliptical galaxies.

\section*{Acknowledgments}
MVM thanks Jochen Liske for useful discussion on quasar absorption line
systems and galactic envelopes.


\begin{thebibliography}{}


\bibitem[1989]{bar89}Barthel P., 1989, ApJ, 336, 606

\bibitem[1990]{baum90} Baum S., Heckman T., van Breugel W., 1990, ApJS, 74, 389

\bibitem[2000]{baum00} Baum S., McCarthy P., 2000, AJ, 119, 2634

\bibitem[2000]{bin2000}Binette L., Kurk J., Villar-Mart\'\i n M.,  R\"ottgering H.,
2000, A\&A, 356, 23

\bibitem[1985]{bor85}B\"ohringer H., 1985, RvMA, 8, 2595

\bibitem[1997]{cari97} Carilli C., R\"ottgering H., van Ojik R., Miley G., van Breugel
W., 1997, ApJS, 109, 1

\bibitem[1999]{carr99} Carroll B, Ostlie D., 1999, in 
{\it An Introduction to Modern Astrophysics}. Addison-Wesley Publishing Company

\bibitem[1990]{cham90} Chambers K., Miley G., van Breugel W., 1990, ApJ, 363, 21 

\bibitem[2001]{chen01} Chen H-W., Lanzetta K., Webb J., 2001, ApJ, 556, 158

\bibitem[1998]{clark98} Clark N., Axon D., Tadhunter C., Robinson C., O'Brien P.,
1998, ApJ, 494, 546

\bibitem[1987]{fab87} Fabian A., Crawford C., Johnstone R., Thomas P., 1987, MNRAS, 963

\bibitem[1995]{goo95}Goodrich R., Cohen M., Putney A., 1995, PASP, 107, 179


\bibitem[2001]{gu01}Gu M., Cao X., Jiang D., 2001, MNRAS, 327, 1111

\bibitem[2001]{hr2001}Haiman Z., Ress M., 2001, ApJ, 556, 87 (HR2001)

\bibitem[1985]{heck85}Heckman T.,  Illingworth G., Miley G., van Breugel W., 1985,
 ApJ, 299, 41

\bibitem[1990]{heck89}Heckman T., Baum S., van Breugel W., McCarthy P., 1989, ApJ,
338, 48

\bibitem[1990]{heck90}Heckman T., Armus L., Miley G., 1990, ApJS, 74, 833 

\bibitem[1991]{heck91}Heckman T., Lehnert M, van Breugel W., Miley G.,  1991, ApJ,
370, 78

\bibitem[2001]{2001} Kurk J., R\"ottgering H., Miley G.,  Pentericci L., 2001, astro-ph/0102337

\bibitem[1995]{lan95} Lanzetta K., Bowen D., Tytler D., Webb J., 1995, ApJ, 442, 538

\bibitem[1998]{len98} Lehnert M., Becker R., 1998, A\&A, 332, 514

 \bibitem[1990a]{mac90a}  McCarthy P.J., Spinrad H., van Breugel W., Liebert J.,
Dickinson M., Djorgovski S., Eisenhardt P., 1990a, ApJ, 365, 487

\bibitem[1990b]{mac90b} McCarthy P., Spinrad H., Dickinson M., van Breugel W.,
 Liebert J., Djorgovski S., Eisenhardt P., 1990b, ApJ, 365, 487

                   

\bibitem[1996]{mac96}  McCarthy P.J., Baum S., Spinrad H., 1996, ApJS, 106, 281

\bibitem[2000]{mor00} Morganti R., Oosterloo S., Tinti S., Tadhunter C., Wills K., 
van Moorsel G., 2002, astro-ph/0112269

\bibitem[1984]{nul84}Nulsen P., Stuart G.C., Fabian A.,  1984, MNRAS, 208, 185

\bibitem[1995]{oke95} Oke et al., 1995, PASP, 107, 375

\bibitem[2002]{oos02} Oosterloo T., Morganti R., Sadler E., Vergani D., Caldwell N.,
2002, AJ, 123, 729

\bibitem[2001]{pet01}Pettini M., Shapley A.,  Steidel C.,  Cuby J.C., Dickinson M.,
Moorwood A., Adelberger K., Giavalisco M.,  2001, ApJ, 554, 981


\bibitem[1999]{pente99} Pentericci L., McCarthy P., R\"ottgering H., Miley G.,
McCarthy P., Spinrad H.,  van Breugel W., Macchetto F., 1999, A\&A, 341, 329


\bibitem[1976]{ray76} Raymond J., Cox D., Smith B., 1976, ApJ, 204, 290

\bibitem[1995a]{rot95a}R\"ottgering H., Miley G., Chambers K.C., 1995, A\&AS, 114, 51 


\bibitem[1995b]{rot95b}R\"ottgering H., Hunstead R., Miley G., van Ojik R., Wieringa M.,
1995, MNRAS,  277, 389

 
\bibitem[1998]{sa98}Sawicky \& Yee, 1998, AJ, 115, 1329

\bibitem[2001]{ste00} Steidel C., Adelberger K., Shapley A., Pettini M., Dickinson M., 
Giavalisco M., 2000, ApJ, 532, 170


\bibitem[2001]{carmen01}Sol\'orzano-I\~narrea  C., Tadhunter C., Axon D., 2001, MNRAS, 323, 965

\bibitem[1996]{stock96}Stockton A., Ridgway S., Kellogg M., 1996, AJ, 112, 902

\bibitem[1989]{tad89} Tadhunter C., Fosbury R., Quinn P., 1989, MNRAS, 240, 225

\bibitem[2001]{tanig01}Taniguchi Y., Shioya Y., Kakazu Y., 2001, ApJ, 562, 15

\bibitem[1988]{tho88} Thomas P., 1988, MNRAS, 235, 315

\bibitem[1996]{ojik96}van Ojik R., R\"ottgering H., Carilli C.L., Miley G.K., Bremer M.N.,
 Macchetto F., 1996, A\&A, 313, 25
 
\bibitem[1997]{ojik97}van Ojik R., R\"ottgering H., Miley G., Hunstead R., 
1997, A\&A, 317, 358

\bibitem[2001]{vern01}Vernet J., Fosbury R., Villar-Mart\'\i n M.,
 Cohen M., Cimatti, A.; di Serego Alighieri S.,
 Goodrich R., 2001, A\&A, 366, 7

\bibitem[1996]{villar96} Villar-Mart\'\i n M., Binette L., Fosbury R., 1996, 
A\&A, 312, 751 

\bibitem[1997]{villar97} Villar-Mart\'\i n M., Tadhunter C., Clark N., 1997, 
A\&A, 323, 21



\bibitem[1999]{villar99} Villar-Mart\'\i n M., Binette L., Fosbury R., 1999, 
A\&A, 346, 7



\bibitem[1999]{villar99} Villar-Mart\'\i n M., Alonso-Herrero A., di Serego-Alighieri S., Vernet J.,
2001, A\&A, 147, 291

\end{thebibliography}
\end{document}